\documentclass[%
 reprint,
 amsmath,amssymb,
 aps,
]{revtex4-2}
\usepackage{amssymb}
\usepackage{bm}
\usepackage[version=4]{mhchem}
\usepackage{amsfonts}
\usepackage{physics}
\usepackage{float}
\usepackage{caption}
\usepackage{tikz}
\usepackage{mathtools,leftindex,tensor}
\usepackage{abraces}
\usepackage{dsfont}
\usepackage{comment}
\usepackage{graphicx}
\usepackage{dcolumn}

\begin{document}

\title{Flavoured Lattice Schwinger Model with Chiral Anomaly  }

\author{Dogukan Bakircioglu}
\affiliation{%
  \small Universit\'{e} Paris-Saclay, INRIA, CNRS, ENS Paris-Saclay,
  LMF, 91190 Gif-sur-Yvette, France
  \vspace{0.4cm}%
}

  \begin{abstract}
We introduce the \emph{flavoured lattice Schwinger model}, a 
$(1{+}1)$-dimensional $U(1)$ lattice gauge theory in which the 
fermion doubling problem is resolved by staggering a $\mathbb{Z}_{2}$ 
flavour degree of freedom rather than staggering chirality. Unlike 
the standard approaches, this construction preserves an exact axial 
$U(1)$ symmetry at finite lattice spacing. We derive the continuum 
limit, showing that the model reduces to the \emph{two-flavour} 
massless Schwinger model, with flavours $\alpha\in\{0,1\}$ sharing 
one dynamical $U(1)$ gauge field. The central result is a 
well-defined, regularised, gauge-invariant lattice axial charge 
$Q_{G}^{A}$ whose continuum non-conservation 
$\langle dQ_{G}^{A}/dt\rangle = -(2g/\pi)\!\int\! dx\,\langle E(x)\rangle$ 
arises as a direct dynamical consequence of minimal gauge coupling. 
A particle-hole transformation on the $\chi$ flavour exposes a 
hidden $U_{L}(2)\times U_{R}(2)$ chiral symmetry; non-Abelian 
bosonisation then identifies the model with a massive abelian 
Schwinger sector tensored with the level-$1$ $SU(2)$ 
Wess--Zumino--Witten model. Finally, we show that embedding the 
flavoured fermions in a ribbon-shaped $(2{+}1)$D 
Bernevig--Hughes--Zhang topological insulator and gauging the bulk 
in a constant background field factorises the boundary theory into 
\emph{two decoupled} single-flavour Schwinger models, one on each 
edge, identifying the lattice factor of $2$ as one quantum of 
Schwinger anomaly per edge.
\end{abstract}

\maketitle

\section{\label{sec:intro}Introduction}

The Schwinger model, quantum electrodynamics (QED) in $(1{+}1)$-dimensions, is among the most celebrated exactly solvable models in quantum field theory~\cite{Schwinger:1951nm,Schwinger:1962tp}. Despite its apparent simplicity, it exhibits a rich physical structure: the photon acquires a mass via the Schwinger mechanism, fermions are confined, chiral symmetry is anomalously broken, and the vacuum supports a non-trivial $\theta$-angle structure.\\

Over the past decade, there has been increasing activity in the study of the lattice Schwinger model using Hamiltonian (real-time) approaches, including tensor-network and related methods~\cite{Magnifico_2020, Chatterjee_2025,itou2024dmrgstudythetadependentmass,Dempsey:2022nys, ohata2023phasediagramnearquantum, Banuls2013, Banuls:2019bmf, Silvi_2014, 10.21468/SciPostPhysLectNotes.8, PhysRevD.104.114501, Dalmonte02072016}, as well as digital quantum simulation and circuit-based approaches~\cite{Arrighi_2020, Piroli_2020, sellapillay2022discreterelativisticspacetimeformalism, Brun:2025ajz}. This growing interest is largely driven by recent advances in experimental quantum simulation platforms~\cite{PhysRevB.100.115152, PhysRevX.10.021041, PhysRevA.98.032331, Martinez2016, PhysRevLett.109.175302, PhysRevA.88.023617, PhysRevLett.109.125302, PhysRevA.83.033625}.\\

In contrast, the development of fundamentally new strategies to overcome the notorious \emph{fermion doubling} (FD) problem---appearance of non-physical fermionic modes in lattice theory~\cite{latticefermions,wilsonlgt}---has seen comparatively less progress in recent years. In standard lattice formulations, however, exact \emph{onsite} chiral symmetry is either explicitly broken at finite lattice spacing~\cite{wilsonlgt,Wilson-quarks,latticefermions} or replaced by the Ginsparg--Wilson relation~\cite{SHARATCHANDRA1981205,ginsparg-wilson,Kaplan:1992bt,Shamir:1993zy}, and is fully recovered only in the continuum limit. Consequently, a straightforward, gauge-invariant definition of the axial charge at finite lattice spacing is generally lacking. Furthermore, in formulations with explicit breaking, the symmetry is violated directly by the lattice action. This contrasts with the continuum theory, where the classical $U(1)_A$ symmetry is broken strictly anomalously.\\

This is because all known solutions to the fermion doubling (FD) problem proceed by deletion of the doublers. However, due to a global constraint arising from the compactness of the Brillouin zone of the lattice, doublers must carry the opposite chirality relative to the physical solutions~\cite{Nielsen1981,NielsenI1981,NielsenII1981,Friedan:1982nk}. Hence, entirely removing the doublers appears to break chiral symmetry. However, interpreting the doublers as supplying the missing chiral partner is incorrect: doublers carry the same quantum numbers as their physical counterparts, whereas genuinely chirally asymmetric fermions in Nature — such as those coupling to the weak interaction — carry different quantum numbers depending on their chirality ~\cite{Goldhaber:1958nb,Feynman:1958ty,Sudarshan:1958vf}. Since the doublers fail to reproduce this quantum-number asymmetry, they should not be counted as realizing the chiral symmetry observed in Nature ~\cite{Nielsen1981,NielsenI1981,NielsenII1981}. Consequently, if one
wishes to claim that lattice regularisation of QFTs is fundamental, a new interpretation
of fermion doubling must be provided. \\

In this work, we apply a new way of fixing FD, introduced in the
context of quantum cellular automata (QCA)~\cite{bakircioglu-fd-qca}, namely
\emph{flavour-staggering-only}. It introduces a $ \mathbb{Z}_{2}$ flavour degree of freedom that is staggered across lattice sites, we refer to this scheme as \emph{flavoured fermions}. Instead of deleting the doublers, it flavours them. This maps the unphysical solutions into flavoured physical solutions. In contrast with \emph{staggered fermions}~\cite{latticefermions}, continuum physical degrees of freedom are not being staggered, only the flavour degree of freedom is.\\

Interpreting doublers as new flavours may seem to leave the problem unresolved, since the doublers do still remain. However, as we make precise in Sec. \ref{sec:fd}, the problem is not merely the number of solutions the lattice theory admits, but the unphysical nature of the doublers.\\

We begin by applying the flavoured fermion strategy in a 
$(1{+}1)$-dimensional Hamiltonian framework. The resulting lattice 
Hamiltonian preserves \emph{both} vector and axial $U(1)$ symmetries 
at the lattice level for the two flavours combined; in the ungauged 
continuum limit the symmetries factorise into two free Dirac sectors, 
while the gauged theory is the two-flavour massless Schwinger model 
with a single shared dynamical $U(1)$ gauge field.\\

The flavoured construction does not violate the Nielsen-Ninomiya theorem~\cite{Nielsen1981,NielsenI1981,NielsenII1981}, 
but rather circumvents it by relaxing the assumption that doublers must be eliminated. Locality, Hermiticity, and the 
correct continuum limit are all maintained.\\

We systematically study the \emph{flavoured lattice Schwinger model},
obtained by coupling the flavoured fermion lattice to a compact $U(1)$
gauge field.
The main results are:
(i) the dispersion relation and continuum limit yield the 
\emph{two-flavour} massless Schwinger model, with flavours 
$\alpha\in\{0,1\}$ sharing a single dynamical $U(1)$ gauge field;
(ii) a \emph{gauge-invariant} lattice axial charge $Q_{G}^{A}$ is 
constructed and shown to commute with the hopping Hamiltonian up to 
$O(a^{2})$, so that its non-conservation is purely dynamical;
(iii) the chiral anomaly is computed,
\begin{equation*}
    \left\langle\frac{dQ_{G}^{A}}{dt}\right\rangle
    = -\frac{2g}{\pi}\int dx\,\langle E(x)\rangle ,
\end{equation*}
the factor $2$ reflecting the doubled fermionic content;
(iv) the anomaly coefficient is shown to be independent of the gauge 
background;
(v) a particle-hole transformation on the $\chi$ flavour exposes a 
hidden $U_{L}(2)\times U_{R}(2)$ chiral symmetry whose non-Abelian 
bosonisation factorises the low-energy theory into a massive abelian 
Schwinger boson and a decoupled level-$1$ $SU(2)$ Wess--Zumino--Witten 
sector; and
(vi) gauging the $(2{+}1)$D Bernevig--Hughes--Zhang topological 
insulator embedding in a constant background electric field 
factorises the boundary theory into \emph{two decoupled} 
single-flavour Schwinger models---one per edge---each carrying one 
quantum of Schwinger anomaly.\\

Admittedly, this emergent $\mathbb{Z}_2$ flavour symmetry is not a fundamental symmetry of the Standard Model, however.  Therefore, a mechanism that geometrically separates the two $\mathbb{Z}_2$ flavour sectors 
must be provided, such that they are spatially separated.\\ 

We show that one way to obtain this physical separation is by considering a $2+1$-dimensional ribbon shaped topological insulator (TI)~\cite{Qi_2011,konig,Bernevig_2006,bernevig2013topological,Shen2017-za}. Each boundary hosts one flavour in the form of helical edge states, in direct analogy with domain wall fermions, this is a natural geometric realization.\\

The paper is organised as follows.
Section~\ref{sec:lattice-schwinger} introduces the lattice Schwinger
model, its Gauss law, and reviews the FD problem and the staggered
fermion solution.
Section~\ref{sec:flavoured-fermions} develops the flavoured fermion
construction, its symmetries, and its continuum limit.
Section~\ref{sec:chiral anomaly} is devoted to the chirally anomalous
response of the flavoured lattice Schwinger model and its low-energy
bosonisation: Subsec.~\ref{sec:lattice-anomaly} derives the flavoured
lattice Schwinger anomaly, including the gauge-invariant axial charge
$Q_{G}^{A}$, the flavoured Gauss law, and the lattice anomaly
equation; Subsec.~\ref{sec:WZW} applies a particle-hole transformation
on the $\chi$ flavour, exposes $U_{L}(2)\times U_{R}(2)$ chiral
symmetry at the continuum level, shows how the Gauss law of the
transformed flavour comes to resemble that of staggered fermions, and
identifies the level-$1$ $SU(2)$ Wess--Zumino--Witten sector via
non-Abelian bosonisation.
Section~\ref{sec:topological-insulator} then turns to a geometric
realisation of the model: Subsec.~\ref{sec:TI-realisation} embeds the
flavoured fermions in a $2{+}1$D TI and derives the helical edge
spectrum, and Subsec.~\ref{sec:gauged-TI} gauges this embedding in a
uniform background electric field, showing that the boundary theory
factorises into two decoupled edge theories that each reproduce one
quantum of Schwinger anomaly.
We conclude in Sec.~\ref{sec:prospects} with a discussion of the
results.
Appendix~\ref{app:correlators} contains the detailed derivation of
the fermionic correlators entering the anomaly calculation.

\section{\label{sec:lattice-schwinger}Lattice Schwinger Model}

The Hamiltonian of the massless Schwinger model on a lattice with
spacing $a$ can be written as~\cite{hamiltonianlgt}

\begin{equation}\label{eq:Hsch}
\begin{split}
    H_{\text{sch}} =
    \sum_{n\in\mathbb{Z}} \frac{i}{2a}
    \Bigl[(\psi^{+}_{n})^{\dagger} U_{n}^{\dagger}\psi^{+}_{n+1}
         -(\psi^{-}_{n})^{\dagger} U_{n}^{\dagger}\psi^{-}_{n+1}\Bigr]
    + \text{h.c.}\\
    +\;\frac{1}{2}\,g^{2}a\,L_{n}^{2},
\end{split}
\end{equation}

\noindent where $(\psi^{\pm}_{n})^{\dagger}$ and $\psi^{\pm}_{n}$ are
fermionic creation and annihilation operators, $U_{n}=e^{igA_{n}a}$
is the $U(1)$ parallel transporter with gauge coupling $g$ and
one-dimensional gauge field $A_{n}$, and $L_{n}$ is the electric
field operator.
The operators $\psi^{+}$ and $\psi^{-}$ annihilate left- and
right-moving (positive and negative chirality) fermions, respectively.
For these operators, we have the following algebras

\begin{equation}\label{eq:comm-rel}
\begin{split}
    \{\psi^{\pm}_{n},\psi^{\pm}_{m}\}=0, \quad
    \{(\psi^{\pm}_{n})^{\dagger},\psi^{\pm}_{m}\}
    =\delta_{n,m}\,\delta^{\pm,\pm},\\[4pt]
    [L_{m},U_{n}]=\delta_{n,m}\,U_{n}, \quad
    [L_{m},U_{n}^{\dagger}]=-\delta_{n,m}\,U_{n}^{\dagger}.
\end{split}
\end{equation}

\medskip
 $H_{\text{sch}}$ is symmetric under one-site translation
$T:\psi_{n}^{\pm}\to\psi_{n+1}^{\pm}$, and for a constant vector potential $ \forall A_{n} =A$,  $H_{\text{sch}}$ diagonalizes in the momentum space $p\in[-\pi/a,\pi/a]$, and we find its energy eigenvalues to be

\begin{equation}\label{eq:E-sch}
    E^{\pm}(p) = \pm\,\frac{\sin(-pa + g A\,a)}{a}.
\end{equation}

Because space is discretised, the momentum space is the
\emph{Brillouin zone} (BZ), topologically equivalent to a circle
with $\pi/a$ and $-\pi/a$ identified.

\medskip

$H_{\text{sch}}$ possesses vector and axial $U(1)$ global symmetries,
$U_{V}(1):\psi_{n}^{\pm}\to e^{i\theta}\psi_{n}^{\pm}$ and
$U_{A}(1):\psi_{n}^{\pm}\to e^{\pm i\theta}\psi_{n}^{\pm}$, with
conserved charges

\begin{equation}
    Q^{V,A}=\sum_{n\in\mathbb{Z}}
    \bigl[(\psi^{+}_{n})^{\dagger}\psi^{+}_{n}
    \pm(\psi^{-}_{n})^{\dagger}\psi^{-}_{n}\bigr]
\end{equation}

\noindent satisfying

\begin{subequations}
\begin{align}
    [H_{\text{sch}},Q^{V}]=[H_{\text{sch}},Q^{A}]=0,\\
    [Q^{A},Q^{V}]=0.
\end{align}
\end{subequations}

\paragraph*{Gauss law.}
Since $H_{\text{sch}}$ is built entirely from the covariantly dressed
(Wilson-line) hopping terms $(\psi^{\pm}_{n})^{\dagger}U_{n}^{\dagger}\psi^{\pm}_{n+1}$
together with the diagonal electric term $L_{n}^{2}$, it is by
construction invariant under the local (small) gauge transformations
generated by

\begin{equation}\label{eq:gauss-sch}
    G_{n}=L_{n}-L_{n-1}-Q_{n}, \qquad
    Q_{n}=(\psi^{+}_{n})^{\dagger}\psi^{+}_{n}
         +(\psi^{-}_{n})^{\dagger}\psi^{-}_{n}-\mathds{1},
\end{equation}

\noindent so that $[H_{\text{sch}},G_{n}]=0$ for every $n$; here
$Q_{n}$ is the fermionic charge density at site $n$ relative to a
uniform background charge of one unit per site. As usual, $G_{n}$ is
not required to vanish as an operator identity but instead defines
the physical Hilbert space through the constraint

\begin{equation}\label{eq:gauss-sch-phys}
    G_{n}\lvert\text{phys}\rangle=0 \qquad \forall n .
\end{equation}

\noindent This is the ordinary (non-staggered) lattice Gauss law of
Kogut--Susskind type~\cite{hamiltonianlgt}. The background charge in
Eq.~\eqref{eq:gauss-sch} is uniform because both chiralities
$\psi^{+}_{n},\psi^{-}_{n}$ live on \emph{every} site $n$; this should
be contrasted with the staggered fermion Gauss law reviewed in
Sec.~\ref{sec:staggered}, Eq.~\eqref{eq:gauss-stg}, whose background
charge alternates with the site parity because a staggered site
carries only a single fermionic mode.

\subsection{\label{sec:fd}Fermion Doubling}

We briefly review the fermion doubling (FD) problem, first identified
in Refs.~\cite{hamiltonianlgt,latticefermions}; a modern treatment in
the QCA framework is given in~\cite{bakircioglu-fd-qca}.
The essence of FD is the existence of a degeneracy in the dispersion
relation that persists in the continuum limit.
We demonstrate it on the lattice Schwinger model in the gauge
$ \forall A_{n}=0$ ; this corresponds to free lattice fermions in
$1{+}1$D~\cite{latticefermions}, and the analysis carries through
unchanged.

In the continuum limit $a\to 0$, around the neighbourhood
$\mathcal{U}_{0}^{a}=\{p:\lvert p\rvert\ll 1/a\}$,
Eq.~\eqref{eq:E-sch} becomes

\begin{equation}\label{eq:E-schcont}
    E^{\pm}_{\text{sch}}(p)\;\cong\;\mp\,p ,
\end{equation}

\noindent which is the dispersion relation of the massless Dirac
equation in $1{+}1$D.
Due to the periodicity of the sine, the same continuum limit also
occurs around $\pm\pi/a\mp p$ for $p\ll 1/a$.
The union of these points, for $p\ll 1/a$, forms a second open
neighbourhood
$\mathcal{U}_{\pi}^{a}=\{\pm\pi/a\mp p:p\ll 1/a\}$.
In the exact continuum limit $\mathcal{U}_{0}^{a}$ becomes $(-\infty,\infty)$
(finite momenta), which is physical.
By contrast, $\mathcal{U}_{\pi}^{a}$ contains only
infinite-momentum modes and is therefore unphysical; these are the
\emph{doublers}. It is important to emphasize that FD does not merely involve an increase in the number of solutions, but rather the emergence of unphysical ones. Consequently, FD represents a qualitative challenge rather than a quantitative one.\\

In free theory, doublers may be harmless because one can simply
restrict to finite-momentum initial states.
With interactions ($A_{n}\neq 0$ for some $n$), however, doublers can
be excited and corrupt the continuum limit.
For this reason FD had been taken seriously; solutions fall into two
broad categories.
The first modifies the Hamiltonian to remove the
degeneracy~\cite{ginsparg-wilson,wilsonlgt}.
The second modifies the lattice symmetries to shrink the BZ so that
$\mathcal{U}_{\pi}^{a}$ is excluded~\cite{latticefermions,hamiltonianlgt,bakircioglu-fd-qca}.

\subsection{\label{sec:staggered}Staggered Fermions}

We focus on the second category, and specifically on staggered
fermions~\cite{latticefermions,SHARATCHANDRA1981205}, which solve FD
by shrinking the BZ from $[-\pi/a,\pi/a]$ to $[-\pi/2a,\pi/2a]$,
thereby excluding $\mathcal{U}_{\pi}^{a}$.
This is achieved not by changing the energy eigenvalues but by
restricting how lattice sites are assigned to creation and annihilation
operators: $\psi^{+}_{n},(\psi^{+}_{n})^{\dagger}$ are defined only
on even sites, while $\psi^{-}_{n},(\psi^{-}_{n})^{\dagger}$ reside
only on odd sites (\emph{staggerisation}).
Introducing a single set $\phi_{n}$ with
$\phi_{2n}=\psi_{2n}^{+}$ and $\phi_{2n+1}=\psi_{2n+1}^{-}$, the massless
staggered Hamiltonian is

\begin{equation}\label{eq:Hstg}
    H_{\text{stg}}=\sum_{n\in\mathbb{Z}}
    \frac{i}{2a}\,(\phi_{n})^{\dagger}(\phi_{n+1}-\phi_{n-1}),
\end{equation}

\noindent or equivalently, in terms of the original chiral fields,

\begin{equation}\label{eq:H-stg-chiral}
\begin{split}
    H_{\text{stg}}
    =\sum_{n\;\text{even}}\frac{i}{2a}
    \Bigl[(\psi^{+}_{n})^{\dagger}
          (\psi^{-}_{n+1}-\psi^{-}_{n-1})\\
    +(\psi^{-}_{n+1})^{\dagger}
     (\psi^{+}_{n+2}-\psi^{+}_{n})\Bigr].
\end{split}
\end{equation}

\paragraph*{Continuum limit.}
The continuum limit is taken on the \emph{block lattice}
$\tilde{\psi}^{+}(ka)=(2a)^{-1/2}\phi_{2k}$,
$\tilde{\psi}^{-}(ka)=(2a)^{-1/2}\phi_{2k+1}$.
In the block-lattice formalism, $H_{\text{stg}}$ reads

\begin{equation}\label{eq:H-block-stg}
\begin{split}
    H_{\text{stg}}
    =\sum_{n \in \mathbb{Z}}i
    \Bigl[(\tilde{\psi}^{+}(na))^{\dagger}
          (\tilde{\psi}^{-}(na)-\tilde{\psi}^{-}(na-a))\\
    +(\tilde{\psi}^{-}(na))^{\dagger}
     (\tilde{\psi}^{+}(na+a)-\tilde{\psi}^{+}(na))\Bigr].
\end{split}
\end{equation}

\noindent Defining $x\equiv na$ and applying the Taylor expansion,
the continuum limit becomes

\begin{equation}\label{eq:H-block-cont}
\begin{split}
    H_{\text{stg}}^{c}
    =\int dx\,\bigl[
    (\tilde{\psi}^{+}(x))^{\dagger}(i\partial_{x}\tilde{\psi}^{-}(x))
    +(\tilde{\psi}^{-}(x))^{\dagger}(i\partial_{x}\tilde{\psi}^{+}(x))
    \bigr].
\end{split}
\end{equation}

\paragraph*{Dispersion relation.}
The staggerised theory is invariant under the two-site translation
$T^{2}:\psi^{+}_{n},\psi^{-}_{n+1}\to\psi^{+}_{n+2},\psi^{-}_{n+3}$,
giving a BZ $[-\pi/2a,\pi/2a]$ with dispersion relation

\begin{equation}\label{eq:E-stg}
    E^{\pm}_{\text{stg}}(p)=\mp\,\frac{\sin(pa)}{a}.
\end{equation}

\noindent The continuum limit is taken with block lattice spacing
$a'=2a$:

\begin{equation}\label{eq:E-stg-block}
    E^{\pm}_{\text{stg}}(p')
    =\mp\,\frac{2\sin(p'a'/2)}{a'},
    \quad p'\in[-\pi/a',\pi/a'].
\end{equation}

\noindent As $a'\to 0$, only the modes in $\mathcal{U}_{0}^{a'}$
survive --- there are no doublers.

\paragraph*{Global symmetries.}
Staggerisation restores neither the axial symmetry on the lattice
(chirally opposite operators reside on different sites), nor the
lattice axial charge.
The only exact lattice symmetry retained is the vector symmetry:

\begin{equation}
    [H_{\text{stg}},Q^{V}]=0.
\end{equation}

\noindent The axial symmetry is recovered in the continuum limit
through the block-lattice structure, but the absence of a
well-defined lattice axial charge makes the study of the chiral
anomaly in the staggered formulation indirect \cite{Chatterjee_2025}.\\

\paragraph*{Gauss law.}
For later comparison, recall that the Gauss law of the staggered
formulation reads

\begin{equation}\label{eq:gauss-stg}
    L_{n}-L_{n-1} = \phi_{n}^{\dagger}\phi_{n}
    -\tfrac{1}{2}\bigl[1-(-1)^{n}\bigr],
\end{equation}

\noindent i.e.\ the background charge vanishes on even sites and
equals one unit on odd sites~\cite{latticefermions}. This alternating
background is required because each staggered site carries only the
single fermionic mode $\phi_{n}$, in contrast with the two independent
modes $\psi^{+}_{n},\psi^{-}_{n}$ retained at every site of
Sec.~\ref{sec:lattice-schwinger}, Eq.~\eqref{eq:gauss-sch}.

Although staggered fermions break the naive on-site chiral symmetry, they provide 
an elegant framework for the lattice regularization of
Dirac-K\"{a}hler fermions across a fully discrete spacetime, where spinors are 
represented in terms of differential forms \cite{becher1982dirac,banks1982geometric}. 
This geometric formulation is highly advantageous for constructing extended 
symmetry structures, such as exact lattice supersymmetry \cite{catterall2009exact}. 
However, it is crucial to emphasize that this strict geometric correspondence 
holds only when both space and time are discretized. In a fully discrete $(N+1)$-dimensional 
spacetime, a Dirac-K\"{a}hler  fermion naturally yields a multi-flavor structure 
containing $2^{(N+1)/2}$ copies of the Dirac equation if $N$ is odd, and $2^{N/2+1}$ 
copies if $N$ is even. In a continuous-time (Hamiltonian) framework, however, 
the spatial discretization restricts the available degrees of freedom to a 
dimensionally reduced spatial exterior algebra. Consequently, in $1+1$ dimensions, 
continuous-time staggered fermions fall short of capturing the full $2$-flavor 
structure inherent to the spacetime Dirac-K\"{a}hler formulation, yielding 
instead a single physical flavor.

\section{\label{sec:flavoured-fermions}Flavoured Fermions}

Flavoured fermions are rooted in staggered fermions but differ in a
key respect: instead of staggering the \emph{chirality}, one
introduces a $\mathbb{Z}_{2}$ \emph{flavour} degree of freedom and
staggers that.
We define fermionic fields $\chi^{\pm}$ and $\psi^{\pm}$ with
opposite $\mathbb{Z}_{2}$ charges, and assign $\chi^{\pm}$ to even
sites and $\psi^{\pm}$ to odd sites.
The resulting Hamiltonian in the gauge $A_{n}=0$ is

\begin{multline}\label{eq:Hf}
    H_{f}=\sum_{n\;\text{even}}\frac{i}{2a}
    \Bigl[(\chi^{+}_{n})^{\dagger}(\psi^{+}_{n+1}-\psi^{+}_{n-1})\\
    -(\chi^{-}_{n})^{\dagger}(\psi^{-}_{n+1}-\psi^{-}_{n-1})\Bigr]
    +\text{h.c.}
\end{multline}

\paragraph*{Continuum limit.}
As for staggered fermions, we introduce a block lattice before taking
the continuum limit.
Define
$\zeta_{\circ}^{\pm}(ka)=(2a)^{-1/2}\chi^{\pm}_{2k}$,
$\zeta_{\bullet}^{\pm}(ka)=(2a)^{-1/2}\psi^{\pm}_{2k+1}$,
and use the identity

\begin{multline}\label{eq:trick}
      \zeta^{\pm}(ka)-\zeta^{\pm}(ka-a)=\tfrac{1}{2}\bigl[\zeta^{\pm}(ka+a)-\zeta^{\pm}(ka-a)\bigr]\\
    -\tfrac{1}{2}\bigl[\zeta^{\pm}(ka+a)+\zeta^{\pm}(ka-a)
      -2\zeta^{\pm}(ka)\bigr],
\end{multline}

\noindent where $\zeta^{\pm}$ stands for either $\zeta_{\circ}^{\pm}$
or $\zeta_{\bullet}^{\pm}$.
The second bracket on the right-hand side does not
contribute to the continuum limit as it produces only terms with $ O(a^{2})$; we therefore define the
\emph{reduced Hamiltonian}

\begin{equation}\label{eq:HfR}
\begin{split}
    H_{f}^{R}
    =\sum_{n\in\mathbb{Z}}\frac{i}{2}
    \biggl[(\zeta_{\circ}^{+}(na))^{\dagger}
    \bigl(\zeta_{\bullet}^{+}(na+a)-\zeta_{\bullet}^{+}(na-a)\bigr)\\
    -(\zeta_{\circ}^{-}(na))^{\dagger}
    \bigl(\zeta_{\bullet}^{-}(na+a)-\zeta_{\bullet}^{-}(na-a)\bigr)
    \biggr]+\text{h.c.},
\end{split}
\end{equation}

\noindent for which $\lim_{a\to0}H_{f}=\lim_{a\to0}H_{f}^{R}$.
Diagonalising in the flavour basis with
$\tilde{\psi}_{\alpha}^{\pm}(na)
  =2^{-1/2}\bigl[\zeta_{\circ}^{\pm}(na)+(-1)^{\alpha}\zeta_{\bullet}^{\pm}(na)\bigr]$
for $\alpha\in\{0,1\}$, the reduced Hamiltonian becomes

\begin{equation}\label{eq:HfR-diag}
\begin{split}
    H_{f}^{R}
    =\sum_{\alpha\in\{0,1\}}(-1)^{\alpha}
    \sum_{n\in\mathbb{Z}}\frac{i}{2}
    \biggl[(\tilde{\psi}_{\alpha}^{+}(na))^{\dagger}
    \bigl(\tilde{\psi}_{\alpha}^{+}(na+a)
        \\ -\tilde{\psi}_{\alpha}^{+}(na-a)\bigr)
    -(\tilde{\psi}_{\alpha}^{-}(na))^{\dagger}
    \bigl(\tilde{\psi}_{\alpha}^{-}(na+a)
         -\tilde{\psi}_{\alpha}^{-}(na-a)\bigr)
    \biggr].
\end{split}
\end{equation}

\noindent In the continuum limit, the flavoured Hamiltonian becomes

\begin{equation}\label{eq:Hfc}
\begin{split}
    H_{f}^{c}
    =i\sum_{\alpha\in\{0,1\}}(-1)^{\alpha}\int dx\,
    \biggl[(\tilde{\psi}_{\alpha}^{+}(x))^{\dagger}
           \partial_{x}\tilde{\psi}_{\alpha}^{+}(x)\\
    -(\tilde{\psi}_{\alpha}^{-}(x))^{\dagger}
     \partial_{x}\tilde{\psi}_{\alpha}^{-}(x)\biggr].
\end{split}
\end{equation}

\noindent Note that the $\mathbb{Z}_{2}$ flavour symmetry of the
discrete theory survives intact in the continuum.

\paragraph*{Dispersion relation.}
$H_{f}$ is invariant under the two-site translation
$T^{2}:\chi^{\pm}_{n},\psi^{\pm}_{n}\to\chi^{\pm}_{n+2},\psi^{\pm}_{n+2}$,
which induces the BZ $[-\pi/2a,\pi/2a]$.
The energy eigenvalues are

\begin{equation}\label{eq:dispersion-flavoured}
    E^{\pm}_{0}(p)= \mp\,\frac{\sin(pa)}{a}, \qquad
    E^{\pm}_{1}(p)= \pm\,\frac{\sin(pa)}{a}.
\end{equation}

\noindent In the block-lattice continuum limit ($a'=2a$):

\begin{equation}\label{eq:dispersion-flavoured-block}
    E^{\pm}_{0}(p')=\mp\,\frac{2\sin(p'a'/2)}{a'}, \qquad
    E^{\pm}_{1}(p')=\pm\,\frac{2\sin(p'a'/2)}{a'}.
\end{equation}

\noindent There are no doublers in either sector.
However, the Hilbert space is twice as large as that of the
continuum Schwinger model: the $\alpha=1$ sector, with its
opposite-sign dispersion, corresponds precisely to the doubler modes,
now promoted to legitimate physical degrees of freedom.\\

\paragraph*{Conserved quantities.}
$H_{f}$ possesses both $U_{V}(1)$ and $U_{A}(1)$ symmetries: $U_{V}(1):\{\chi_{n}^{\pm},\psi_{n+1}^{\pm}\}  \to\{e^{i\theta}\chi_{n}^{\pm},e^{i\theta}\psi_{n+1}^{\pm}\}$ and $U_{A}(1):\{\chi_{n}^{\pm},\psi_{n+1}^{\pm}\} \to\{e^{\pm i\theta}\chi_{n}^{\pm},e^{\pm i\theta}\psi_{n+1}^{\pm}\}$. The corresponding  conserved charges are

\begin{equation}\label{eq:QfVA}
\begin{split}
    Q^{V,A}_{f}
    = \frac{1}{2}\sum_{n\;\text{even}}
    \Bigl[(\chi^{+}_{n})^{\dagger}\chi^{+}_{n}
         \pm(\chi^{-}_{n})^{\dagger}\chi^{-}_{n}\\
    +(\psi^{+}_{n+1})^{\dagger}\psi^{+}_{n+1}
    \pm(\psi^{-}_{n+1})^{\dagger}\psi^{-}_{n+1}\Bigr],
\end{split}
\end{equation}

\noindent with $[H_{f},Q_{f}^{V}]=[H_{f},Q_{f}^{A}]=0$ and
$[Q_{f}^{A},Q_{f}^{V}]=0$.\\

The exact lattice $U_{A}(1)$ symmetry exhibited by $H_f$ should be understood as a symmetry acting on the full flavoured Hilbert space, rather than on a single chiral sector. In particular, while the total axial charge $Q_A^f$ is conserved on the lattice, charges associated with individual flavour sectors are not separately conserved.\\

We also have the following conserved charges in the continuum,
\begin{equation}
    \tilde{Q}_{0,1}^{V,A}
    =\int dx\bigl[(\tilde{\psi}_{0,1}^{+})^{\dagger}\tilde{\psi}_{0,1}^{+}
    \pm(\tilde{\psi}_{0,1}^{-})^{\dagger}\tilde{\psi}_{0,1}^{-}\bigr],
\end{equation}

\noindent and the lattice charges flow to

\begin{equation}
    \lim_{a'\to0}Q_{f}^{V,A}
    =\bigl(\tilde{Q}_{0}^{V,A}+\tilde{Q}_{1}^{V,A}\bigr).
\end{equation}

As noted in Ref.~\cite{Schwinger:1962tp}, the expectation values
$\langle Q_{f}^{V}\rangle$ and $\langle Q_{f}^{A}\rangle$ diverge as
$a\to 0$, necessitating regularisation.
We introduce the \emph{regularised charges}

\begin{equation}\label{eq:QR}
\begin{split}
    Q^{V,A}_{R}
    = \frac{1}{2}\sum_{n\;\text{even}}
    \Bigl[(\chi^{+}_{n})^{\dagger}\psi^{+}_{n+1}
         \pm(\chi^{-}_{n})^{\dagger}\psi^{-}_{n+1}\\
    +(\psi^{+}_{n+1})^{\dagger}\chi^{+}_{n}
    \pm(\psi^{-}_{n+1})^{\dagger}\chi^{-}_{n}\Bigr],
\end{split}
\end{equation}

\noindent which flow to

\begin{equation}
    \lim_{a'\to0}Q_{R}^{V,A}
    =\bigl(\tilde{Q}_{0}^{V,A}-\tilde{Q}_{1}^{V,A}\bigr).
\end{equation}

Even though  $\tilde{Q}_{0}^{V,A} -  \tilde{Q}_{1}^{V,A} $ is conserved in the continuum, $ Q_{R}^{V,A}$ are not conserved on the lattice.  This is indeed due to the second order terms that we neglected at Eq.\eqref{eq:trick}, and  we have the following commutation relations,
\begin{equation}
    \begin{split}
        [H_{f},Q_{R}^{V,A}] = \frac{i}{2a}\sum_{n \, \, \textit{even}} (\chi_{n}^{+})^{\dagger} ( 2 \chi_{n}^{+}- \chi_{n-2}^{+}-\chi_{n+2}^{+})\\
        - (\psi_{n+1}^{+})^{\dagger} ( 2 \psi_{n+1}^{+}- \psi_{n-1}^{+}-\psi_{n+3}^{+})\\ \pm (\chi_{n}^{-})^{\dagger} ( 2 \chi_{n}^{-}- \chi_{n-2}^{-}-\chi_{n+2}^{-})\\
        \mp (\psi_{n+1}^{-})^{\dagger} ( 2 \psi_{n+1}^{-}- \psi_{n-1}^{-}-\psi_{n+3}^{-}).
    \end{split}
\end{equation}
In the continuum limit, these terms cancel, and we obtain $ \lim_{a \to 0}[H_{f},Q_{R}^{V,A}]= 0 $.

\paragraph*{The no-go theorem.}
From Eq.~\eqref{eq:dispersion-flavoured}, the energy eigenvalues of flavour $\alpha=0$ and $\alpha=1$ have opposite orientations: while $\tilde{\psi}_{0}^{\pm}$ carries chiral charge $\pm$, $\tilde{\psi}_{1}^{\pm}$ carries a chiral charge $\mp$. This sign reversal is a direct consequence of the doublers residing at $\mathcal{U}_{1}^{a}$ rather than $\mathcal{U}_{0}^{a}$, and reflects a fundamental constraint imposed by the compactness of the Brillouin zone. This is indeed due to an application of the Atiyah--Singer index theorem~\cite{Atiyah:1963zz, Eguchi:1980jx}, which states that the net number of chiral zero modes of the Dirac operator must be equal to the Euler characteristic of the compact surface on which it is defined.  A closely related argument was given in Ref.~\cite{NielsenI1981}, where it was concluded that the weak interaction cannot be realised on the lattice, since the weak interaction couples differently to fermions of opposite chirality\footnote{For instance, left-handed fermions carry nonzero weak isospin, whereas right-handed fermions are isospin singlets.}. However, we stress that the index theorem does \emph{not} forbid chiral asymmetry in general; it only forces a fermion and its doubler to carry opposite chirality. Consequently, in the staggered fermion approach to the doubling problem---where $\psi_{n}^{-}$ is identified as the doubler of $\psi_{n}^{+}$---encoding the weak interaction on the lattice is impossible. By contrast, the flavoured fermion formulation introduces additional degrees of freedom that, in principle, admit a chiral-asymmetric coupling \cite{bakircioglu-fd-qca}. The price to pay, however, is the emergence of a $\mathbb{Z}_{2}$ flavour symmetry in the continuum limit that is absent from the Standard Model. A mechanism must therefore be provided that geometrically distinguishes 
fermions of opposite $\mathbb{Z}_2$ flavour, rendering them physically 
separable without requiring the symmetry to be explicitly broken.
\section{\label{sec:chiral anomaly}%
  Chiral anomaly and low-energy bosonisation}

This section analyses the chirally anomalous response of the flavoured
lattice Schwinger model and its equivalent low-energy identification via
non-Abelian bosonisation. We first construct the gauge-invariant lattice
axial charge and derive the lattice anomaly equation
(Subsec.~\ref{sec:lattice-anomaly}), and then describe the $SU(2)$
Wess--Zumino--Witten sector exposed by a particle-hole transformation
(Subsec.~\ref{sec:WZW}). A geometric interpretation of the anomaly, in
terms of a ribbon topological insulator embedding whose gauging
reproduces one quantum of Schwinger anomaly per edge, is developed
separately in Sec.~\ref{sec:topological-insulator}.

\subsection{\label{sec:lattice-anomaly}%
  Chiral anomaly in the flavoured lattice Schwinger model}

In this subsection we couple the flavoured fermions to the $U(1)$ gauge 
field and study the chiral anomaly.

\paragraph*{Flavoured Schwinger model.}
When the flavoured fermion solution is applied to $H_{\text{sch}}$,
the result is

\begin{equation}\label{eq:Hschf}
\begin{split}
    H_{\text{sch}}^{f}
    =\sum_{n\;\text{even}}\frac{i}{2a}
    \Bigl[(\chi^{+}_{n})^{\dagger}(U_{n}^{\dagger}\psi^{+}_{n+1}
          -U_{n-1}\psi^{+}_{n-1})\\
    -(\chi^{-}_{n})^{\dagger}(U_{n}^{\dagger}\psi^{-}_{n+1}
      -U_{n-1}\psi^{-}_{n-1})\Bigr]\\
    +\,\frac{1}{2}g^{2}a(L_{n}^{2}+L_{n+1}^{2})
    +\text{h.c.}
\end{split}
\end{equation}

\noindent We call this the \emph{flavoured lattice Schwinger model}.
Its continuum limit is taken using the same block-lattice recipe as
in Sec.~\ref{sec:flavoured-fermions}.
We introduce $\tilde{U}(ka)=U_{2k+1}$, $\tilde{A}(ka)=A_{2k+1}$,
and $\tilde{E}(ka)=gL_{2k+1}$ for $k\in\mathbb{Z}$ (inter-block
gauge links).
Using the generalised block-lattice identity

\begin{equation}\label{eq:trick-gauge}
\begin{split}
    &\tilde{U}^{\dagger}\!\bigl(ka-\tfrac{a}{2}\bigr)\zeta^{\pm}(ka)
    -\tilde{U}(ka-a)\zeta^{\pm}(ka-a)\\
    &=\tfrac{1}{2}\bigl[\tilde{U}^{\dagger}(ka)\zeta^{\pm}(ka+a)
      -\tilde{U}(ka-a)\zeta^{\pm}(ka-a)\bigr]\\
    &\quad+\bigl[\tilde{U}^{\dagger}\!\bigl(ka-\tfrac{a}{2}\bigr)-1\bigr]
      \zeta^{\pm}(ka)+O(a^{2}),
\end{split}
\end{equation}

\noindent and performing the flavour diagonalisation of
Sec.~\ref{sec:flavoured-fermions}, we obtain

\begin{equation}\label{eq:Hschf-cont}
\begin{split}
    \lim_{a\to0}H_{\text{sch}}^{f}
    =i\sum_{\alpha\in\{0,1\}}(-1)^{\alpha}\int dx\,
    \biggl[(\tilde{\psi}_{\alpha}^{+})^{\dagger}
           (\partial_{x}-ig\tilde{A})\tilde{\psi}_{\alpha}^{+}\\
    -(\tilde{\psi}_{\alpha}^{-})^{\dagger}
     (\partial_{x}-ig\tilde{A})\tilde{\psi}_{\alpha}^{-}\biggr]
    +\frac{1}{2}\int dx\,\tilde{E}^{2}(x),
\end{split}
\end{equation}

\noindent where $\tilde{A}(x)=\lim_{a\to0}\tilde{A}(ka)$ and
$\tilde{E}(x)=\lim_{a\to0}\tilde{E}(ka)$.

\paragraph*{Physical content of the continuum limit.}
The Hamiltonian~\eqref{eq:Hschf-cont} is the \emph{two-flavour 
massless Schwinger model}: the two flavour sectors $\alpha\in\{0,1\}$ 
propagate as independent Dirac species coupled to a \emph{common} 
dynamical $U(1)$ gauge field $\tilde A(x)$, with the shared kinetic 
term $\tfrac{1}{2}\!\int\! dx\,\tilde E^{2}(x)$. The flavours interact 
through the exchange of the Schwinger photon, in close analogy with 
two flavours of quarks sharing one gluon. The Schwinger mechanism 
then generates a single massive iso-singlet boson, with mass enhanced 
by the doubled content~\cite{Schwinger:1962tp},
\begin{equation}\label{eq:Schwinger-mass}
    m^{2}=\frac{N_{f}\,g^{2}}{\pi}=\frac{2g^{2}}{\pi},
    \qquad N_{f}=2,
\end{equation}
while the lattice $\mathbb{Z}_{2}$ flavour symmetry is exact. The 
lattice axial charge $Q_{G}^{A}$ of Eq.~\eqref{eq:QGA} measures the 
iso-singlet axial charge in the continuum limit, which is the 
combination that carries the Schwinger anomaly. The factor of $2$ in 
Eq.~\eqref{eq:lattice-anomaly} thus reflects the doubled fermionic 
content; it is \emph{not} a sum of two independent anomalies. Its 
geometric reinterpretation as one quantum of Schwinger anomaly per 
edge requires the $(2{+}1)$D topological insulator embedding 
developed later in Sec.~\ref{sec:topological-insulator} 
(Subsec.~\ref{sec:gauged-TI}).

\paragraph*{Gauss law.}
The flavoured lattice Hamiltonian $H_{\text{sch}}^{f}$ is likewise
built entirely from covariantly dressed hopping terms, so it commutes
with the local generators

\begin{equation}\label{eq:gauss-flav}
\begin{split}
    G_{n}=L_{n}-L_{n-1}-\rho_{n}, \\
    \rho_{n}=
    \begin{cases}
    (\chi^{+}_{n})^{\dagger}\chi^{+}_{n}+(\chi^{-}_{n})^{\dagger}\chi^{-}_{n}-\mathds{1}, & n\ \text{even},\\[6pt]
    (\psi^{+}_{n})^{\dagger}\psi^{+}_{n}+(\psi^{-}_{n})^{\dagger}\psi^{-}_{n}-\mathds{1}, & n\ \text{odd},
    \end{cases}
    \end{split}
\end{equation}

\noindent i.e.\ $[H_{\text{sch}}^{f},G_{n}]=0$, and the physical
Hilbert space is again selected by $G_{n}\lvert\text{phys}\rangle=0$
for every $n$. Although the flavour degree of freedom is staggered
across even and odd sites, the Gauss law itself keeps the
\emph{usual}, non-staggered form of Eq.~\eqref{eq:gauss-sch}: both
chiralities of whichever flavour field occupies site $n$
($\chi^{\pm}_{n}$ on even sites, $\psi^{\pm}_{n}$ on odd sites)
contribute to $\rho_{n}$ with the same uniform background charge of
one unit per site. This is a direct consequence of staggering the
\emph{flavour} rather than the \emph{chirality}: every site continues
to host a full chirality pair, so --- unlike genuine staggered
fermions, Eq.~\eqref{eq:gauss-stg} --- no alternating background
charge is required. We note for later use that the electric field
$E(x)$ entering the chiral anomaly equation~\eqref{eq:lattice-anomaly}
is precisely the continuum limit of the field constrained by
Eq.~\eqref{eq:gauss-flav}, sourced by this physical fermionic charge
density.

\paragraph*{Gauge-invariant charges.}
By arguments analogous to those in Sec.~\ref{sec:flavoured-fermions},
$H_{\text{sch}}^{f}$ commutes with the total vector and axial
charges $[Q_{f}^{V,A},H_{\text{sch}}^{f}]=0$.
However, as established in
Ref.~\cite{Schwinger:1962tp}, the expectation
values $\langle Q_{f}^{V,A}\rangle$ diverge in the continuum limit.
The regularised charges $Q_{R}^{V,A}$ of Eq.~\eqref{eq:QR} are
not gauge-invariant.
Their gauge-invariant extension is

\begin{equation}\label{eq:QGA}
\begin{split}
    Q^{V,A}_{G}
    =\sum_{n\;\text{even}}
    \Bigl[(\chi^{+}_{n})^{\dagger}U_{n}^{\dagger}\psi^{+}_{n+1}
         \pm(\chi^{-}_{n})^{\dagger}U_{n}^{\dagger}\psi^{-}_{n+1}\\
    +(\psi^{+}_{n+1})^{\dagger}U_{n}\chi^{+}_{n}
    \pm(\psi^{-}_{n+1})^{\dagger}U_{n}\chi^{-}_{n}\Bigr].
\end{split}
\end{equation}

\noindent A remark on uniqueness: $Q_{G}^{A}$ is the unique
gauge-invariant extension of $Q_{R}^{A}$ at $O(a^{0})$.
Any alternative gauge-invariant bilinear at this order would
differ by terms proportional to additional lattice derivatives or
longer Wilson lines, which are of $O(a)$ or higher and therefore
vanish in the continuum limit without affecting the anomaly
coefficient.
This uniqueness is important: it means that the factor of $2$ in
Eq.~\eqref{eq:lattice-anomaly} is not an artifact of a particular
operator choice but is intrinsic to the flavoured Hilbert space.

\paragraph*{Chiral anomaly.}
We compute $[H_{\text{sch}}^{f},Q_{G}^{V,A}]$.
The hopping part of $H_{\text{sch}}^{f}$ contributes only at
$O(a^{2})$; at leading order

\begin{equation}\label{eq:comm-QG}
    [H_{\text{sch}}^{f},Q_{G}^{V,A}]
    =\frac{1}{4}g^{2}a\sum_{n\;\text{even}}
    [L_{n}^{2},Q_{G}^{V,A}]+O(a^{2}).
\end{equation}

\noindent Using the commutation relations \eqref{eq:comm-rel} and
expanding $[L_{n}^{2},Q_{G}^{V,A}]$, we find

\begin{equation}\label{eq:comm-L2-QG}
\begin{split}
    [H_{\text{sch}}^{f},Q_{G}^{V,A}]= 
     \frac{1}{2}g^{2}a\sum_{n\;\text{even}}L_{n} \hspace{20mm}\\
    \cross \Bigl[(\psi^{+}_{n+1})^{\dagger}U_{n}\chi^{+}_{n}
        \pm(\psi^{-}_{n+1})^{\dagger}U_{n}\chi^{-}_{n} \hspace{20mm}\\
    -(\chi^{+}_{n})^{\dagger}U_{n}^{\dagger}\psi^{+}_{n+1}
    \mp(\chi^{-}_{n})^{\dagger}U_{n}^{\dagger}\psi^{-}_{n+1}\Bigr] + Q_{G}^{V,A}  .
\end{split}
\end{equation}

We now consider a uniform $ U(1)$ background $ \forall U_{n}=e^{ig A a}$ on all links, and for this scenario, the expectation value of the gauge-invariant fermionic bilinear is computed in Appendix~\ref{app:correlators}:
\begin{equation}\label{eq:correlator-main}
    \bigl\langle(\psi^{\pm}_{n+1})^{\dagger} U_{n}\chi^{\pm}_{n}
   \bigr\rangle
    =\mp\frac{i}{\pi}.
\end{equation}
Since $  \bigl\langle(\psi^{\pm}_{n+1})^{\dagger} U_{n}\chi^{\pm}_{n}
   \bigr\rangle$ is pure imaginary, we find $ \langle Q_{G}^{V,A}\rangle= 0$. Inserting Eq.~\eqref{eq:correlator-main} and taking
$a\to0$:
\begin{subequations}\label{eq:anomaly-eqs}
\begin{align}
    \lim_{a\to0}\langle[H_{\text{sch}}^{f},Q_{G}^{V}]\rangle &=0,
    \label{eq:anomaly-vector}\\
    \lim_{a\to0}\langle[H_{\text{sch}}^{f},Q_{G}^{A}]\rangle
    &=\frac{2ig}{\pi}\int dx\,E(x).
    \label{eq:anomaly-axial}
\end{align}
\end{subequations}

\noindent Via $d\langle Q_{G}^{A}\rangle/dt
=i\langle[H_{\text{sch}}^{f},Q_{G}^{A}]\rangle$, the
\emph{lattice chiral anomaly equation} reads

\begin{equation}\label{eq:lattice-anomaly}
    \left\langle\frac{dQ_{G}^{A}}{dt}\right\rangle
    =-\frac{2g}{\pi}\int dx\,\langle E(x)\rangle .
\end{equation}

\noindent Unlike the staggered fermion formulation, where the axial
symmetry is explicitly broken by the staggerisation procedure and no
well-defined lattice axial charge exists, $Q_{G}^{A}$ is a
\emph{gauge-invariant, well-defined lattice observable} that commutes
with the hopping part of $H_{\text{sch}}^{f}$ up to $O(a^{2})$
corrections.
Its non-conservation is therefore entirely dynamical and arises
purely from the minimal coupling to the gauge field.

The factor of $2$ relative to the single-flavour continuum result
$dQ^{A}/dt=(g/\pi)\int E\,dx$~\cite{Schwinger:1962tp}
reflects the doubled fermionic degrees of freedom in the flavoured
construction: the $\pm$ chirality sectors each carry an independent
contribution to the anomaly, and they add coherently in $Q_{G}^{A}$.
The vector charge $Q_{G}^{V}$ remains conserved in the continuum
limit, consistently with the absence of a vector anomaly.
A detailed derivation of the correlator
Eq.~\eqref{eq:correlator-main} is given in
Appendix~\ref{app:correlators}.\\

\emph{Gauge-independence.}
The correlator $\langle(\psi_{n+1}^{\pm})^{\dagger}U_{n}\chi_{n}^{\pm}\rangle$
is computed exactly (Appendix~\ref{app:correlators}) and shown to be
independent of a uniform background field $U_{n}=e^{i\theta}$.
This gauge-independence is a structural feature of $(1{+}1)$D
electrodynamics: the gauge field carries no local degrees of freedom
and can be removed by a gauge transformation up to global holonomy
effects~\cite{Smit_2002}.
Hence the gauge-invariant correlator equals its free-theory value
$\pm i/\pi$ for any background, and the anomaly coefficient is
insensitive to smooth deformations of the gauge field.

\subsection{\label{sec:WZW}Particle-hole transformation and the Wess--Zumino--Witten sector}

The opposite-sign dispersion of the two flavour sectors in Eq.~\eqref{eq:Hfc} indicates that the $\alpha=1$ doubler acts as the particle-hole conjugate of an ordinary Dirac fermion. Undoing this conjugation exposes a hidden $U_L(2)\times U_R(2)$ continuum chiral symmetry. 

Specifically, defining the partial charge conjugation on the $\chi$ sublattice $\mathcal{C}_\chi: \chi^\sigma_n \mapsto (\chi^{-\sigma}_n)^\dagger$ corresponds, in the continuum, to the redefinition of the inverted-dispersion sector:
\begin{equation}\label{eq:PH-alpha}
   \hat\psi^{\sigma}_{0}\equiv\tilde\psi^{\sigma}_{0},\qquad
   \hat\psi^{\pm}_{1}\equiv (\tilde\psi^{\mp}_{1})^{\dagger}.
\end{equation}
This removes the relative sign of the kinetic terms in Eq.~\eqref{eq:Hfc}, mapping the gauged continuum Hamiltonian Eq.~\eqref{eq:Hschf-cont} directly to the standard two-flavour Schwinger model:
\begin{equation}\label{eq:H-cont-U2}
   \hat H_{f}^{c}
   =i\!\sum_{\alpha=0,1}\!\int dx\,\Bigl[(\hat\psi^{+}_{\alpha})^{\dagger}D_{x}\hat\psi^{+}_{\alpha}
   -(\hat\psi^{-}_{\alpha})^{\dagger}D_{x}\hat\psi^{-}_{\alpha}\Bigr]
   +\tfrac{1}{2}\!\int\! dx\,\tilde E^{2},
\end{equation}
where $D_{x}=\partial_{x}-ig\tilde A$. The free part of this theory possesses an explicit $U_L(2)\times U_R(2)$ chiral symmetry acting on the doublet $\Psi^\sigma \equiv (\hat\psi^\sigma_0, \hat\psi^\sigma_1)^T$.

\paragraph*{Gauss law under charge conjugation.}
It is instructive to see how the local constraint
Eq.~\eqref{eq:gauss-flav} is realised in terms of the transformed
field $\hat\chi_{n}^{\sigma}\equiv\mathcal{C}_{\chi}(\chi_{n}^{\sigma})=(\chi_{n}^{-\sigma})^{\dagger}$
underlying the redefinition~\eqref{eq:PH-alpha}. Inverting,
$\chi_{n}^{+}=(\hat\chi_{n}^{-})^{\dagger}$ and
$\chi_{n}^{-}=(\hat\chi_{n}^{+})^{\dagger}$, so the canonical
anticommutators give
$(\chi_{n}^{+})^{\dagger}\chi_{n}^{+}+(\chi_{n}^{-})^{\dagger}\chi_{n}^{-}
=2\mathds{1}-\bigl[(\hat\chi^{+}_{n})^{\dagger}\hat\chi^{+}_{n}
+(\hat\chi^{-}_{n})^{\dagger}\hat\chi^{-}_{n}\bigr]$, and hence the
even-site charge density of Eq.~\eqref{eq:gauss-flav} becomes

\begin{equation}\label{eq:gauss-conjugated}
    \rho_{n}\big|_{n\ \text{even}} = -\hat\rho_{n}, \qquad
    \hat\rho_{n}\equiv(\hat\chi^{+}_{n})^{\dagger}\hat\chi^{+}_{n}
    +(\hat\chi^{-}_{n})^{\dagger}\hat\chi^{-}_{n}-\mathds{1},
\end{equation}

\noindent while the odd-site ($\psi$) contribution is untouched by
$\mathcal{C}_{\chi}$. The background charge on the transformed
sublattice therefore flips sign, $+\mathds{1}\to-\mathds{1}$: in the
hatted variables that make the kinetic term of
Eq.~\eqref{eq:H-cont-U2} manifestly uniform across $\alpha=0,1$, the
$\alpha=1$ flavour is Gauss-law--conjugate to an ordinary fermion,
i.e.\ it is described as a \emph{hole} rather than a particle. This is
structurally the same particle-hole bookkeeping responsible for the
alternating background charge of the genuine staggered fermion Gauss
law, Eq.~\eqref{eq:gauss-stg}: one sublattice is populated by
particles and the complementary one effectively by holes. Here the
distinction is by \emph{flavour} rather than by \emph{site parity}
--- consistent with the flavoured construction staggering flavour
rather than chirality throughout --- but the underlying mechanism is
the same one staggered fermions use.

By applying standard non-Abelian bosonisation~\cite{Witten:1983ar} to this $U_V(1)$-gauged theory, the dynamics factorise. The gauge field couples exclusively to the diagonal $U_V(1)$ current, yielding an anomalous mass $m_\phi^2 = 2g^2/\pi$ for the abelian boson $\phi$. The non-Abelian $SU(2)$ currents remain perfectly decoupled. The low-energy physics of the flavoured Schwinger model thus cleanly decomposes into a massive abelian Schwinger boson and a gapless, level-$1$ $SU(2)$ Wess--Zumino--Witten (WZW) spectator sector:
\begin{equation}\label{eq:LE-decomp}
   H_{\mathrm{sch}}^{f}\;\xleftrightarrow{\;\mathcal{C}_{\chi}\,+\,\text{boson.}\;}\;
   \underbrace{H_{\phi}\bigl(m_{\phi}^{2}=2g^{2}/\pi\bigr)}_{\text{Schwinger sector}}
   \;\oplus\;
   \underbrace{H_{SU(2)_{1}}^{\mathrm{WZW}}}_{\text{Decoupled WZW sector}}.
\end{equation}

In this framework, the doubled fermionic content introduced to resolve fermion doubling simply reorganises at low energies into this neutral, spectator $SU(2)_1$ current algebra, ensuring the exact topological matching required by the overall anomaly.

\section{\label{sec:topological-insulator}%
  Topological Insulator Embedding}

This section develops a geometric realisation of the flavoured lattice
Schwinger model and of the chiral anomaly derived in
Sec.~\ref{sec:chiral anomaly}. Subsec.~\ref{sec:TI-realisation} embeds
the flavoured fermions in a $(2{+}1)$D topological insulator and derives
the helical edge spectrum; Subsec.~\ref{sec:gauged-TI} gauges this
embedding in a uniform background electric field and shows that the
factor of $2$ in the lattice anomaly equation~\eqref{eq:lattice-anomaly}
is realised geometrically as one quantum of Schwinger anomaly per edge.

\subsection{\label{sec:TI-realisation}Realisation as a topological insulator}

In this subsection, we discuss a physical realisation of the emergent 
$\mathbb{Z}_{2}$ symmetry appearing in the continuum limit of the 
flavoured fermions. We embed the $1+1$-dimensional flavoured fermion 
theory into a $2+1$-dimensional bulk modelled as a quantum spin Hall 
topological insulator in ribbon geometry, drawing inspiration from the 
domain wall fermion approach to chiral symmetry on the lattice. We show 
that the two $\mathbb{Z}_{2}$ flavours are naturally realised as helical 
edge states localised on opposite boundaries of the ribbon, providing a 
physical mechanism that spatially separates fermions of different flavour  and thereby resolves the tension identified at the end of Sec.~\ref{sec:flavoured-fermions}.\\

\paragraph*{Domain wall fermions.}  We draw some inspiration from another solution of FD, that is called ''domain wall fermions'' (DWFs) \cite{Kaplan:1992bt,Shamir:1993zy}. In this particular solution, they embed the $4$D theory into a $5$D bulk. For $ s$ being the coordinate of the fifth dimension and $  2 l_{5}$ being the length, they consider a $s$ dependent mass term $M(s)$ that satisfies $\lim_{s \to \pm\infty} M(s) = \pm M_0$. The bulk-boundary correspondence dictates that the topological index of the $5$D Dirac operator, arising from the transition of the bulk mass, requires the existence of zero mode solutions to the Dirac equation. These modes are exponentially localized at the 4D boundaries (the domain walls).  By physically separating the chiral flavours along the fifth dimension, the $U(1)_A$ symmetry is recovered in the $l_5 \to \infty$ limit, as the overlap between the left-handed mode at $s=- l_5$ and the right-handed mode at $s=l_5$ vanishes, allowing for a lattice realization of chiral fermions that satisfies the Ginsparg-Wilson relation. They avoid the Nielsen-Ninomiya theorem by explicitly breaking the chiral symmetry on the bulk theory and recovering it on the boundary as $ l_{5} \to \infty$.\\

\begin{figure}[t]
\centering
\begin{tikzpicture}[scale=0.9]

\fill[gray!15] (0,0) rectangle (7,1.8);
\draw[thick] (0,0) rectangle (7,1.8);

\draw[->, thick, blue] (0.4,1.65) -- (6.6,1.65);
\draw[<-, thick, red]  (0.4,1.45) -- (6.6,1.45);

\draw[<-, thick, blue] (0.4,0.35) -- (6.6,0.35);
\draw[->, thick, red]  (0.4,0.15) -- (6.6,0.15);

\draw[->, blue, thick] (3.5,2.05) -- (3.5,2.35);

\draw[->, red, thick] (3.7,2.35) -- (3.7,2.05);

\node at (3.5,0.9) {\small Topological Insulator};
\node at (0.6,2.5) {\small Vacuum};

\node[left] at (-0.15,0.9) { $ 2l_y$};
\node[below] at (3.5,-0.1) {$l_x \simeq \infty$};

\end{tikzpicture}
\caption{Ribbon geometry with helical edge states. Arrows along the edges indicate propagation direction, while arrows outside the ribbon denote the associated spin degrees of freedom. }
\label{fig:qshe}
\end{figure}
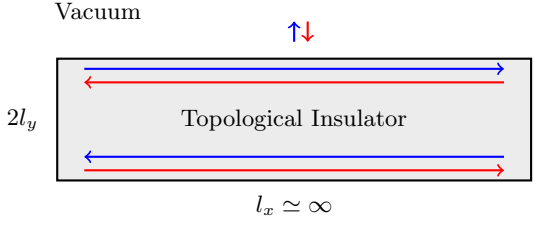

 \paragraph*{Quantum spin Hall effect.}
Following the domain wall fermion approach, we embed the flavoured fermion 
theory into a $2+1$-dimensional spacetime with spatial geometry of a ribbon: 
$x \in (-\infty, \infty)$ and $y \in [-l_y, l_y]$. The bulk is modelled as a 
topological insulator (TI), whose gapped bulk spectrum admits gapless edge 
states that realise free fermions at low energies~\cite{bernevig2013topological}. 
This is precisely the structure required to spatially separate the two flavour 
sectors: a ribbon-geometry TI supports one edge state on each boundary, with 
the two edge modes propagating in opposite directions. We restrict to 
time-reversal-invariant topological insulators, which exhibit the quantum spin 
Hall effect (QSHE)~\cite{Kane_2005, Bernevig_2006, konig}: at each edge, two counter-propagating modes with opposite spin are present (here "spin" is the real electron spin of the QSHE literature; below we identify it with the chirality index ± of our construction, not with the Z2 flavour). The ribbon geometry 
therefore supports four gapless edge modes in total — two at $y = l_y$ and two 
at $y = -l_y$ — as illustrated in Fig.~\ref{fig:qshe}. These are the helical 
edge states on which the two $\mathbb{Z}_2$ flavours will be localised.

 \paragraph*{Our model.} Our aim is to construct a Hamiltonian out of the fermionic fields, $ \chi^{\pm}, \psi^{\pm}$ in the bulk and obtain edge states on the different boundaries having different flavours. We want to continue to the use of square lattice with flavours. For this reason, we study a Hamiltonian in $ 2+1$D that is similar to the two dimensional Bernevig-Hughes-Zhang (BHZ) model \cite{Bernevig_2006,Qi_2011,konig}, where they have a Hamiltonian on a square lattice with orbital degrees of freedom on each lattice point. In BHZ model, there are four degrees of freedom for electrons, they are either at $s-$orbitals or $p-$orbitals with spin up and down. Whereas in our case we do not have spin degrees of freedom nor orbitals, instead we have flavour symmetry and chirality, and we map the flavour degree of freedom to be orbital and the chirality to be spin. We have increased the number of spatial dimensions, we again take them to be discretised, and we keep the number of flavours the same. We have two sub-lattices that are covering the square two-dimensional lattice with the new two-dimensional lattice vectors $a_{1}= a(\hat{x} + \hat{y})$ and $ a_{2} = a(\hat{x} - \hat{y})$, for $ \hat{x}$ and $ \hat{y}$ are the unit vectors, again $ a$ is the lattice separation. In the square two-dimensional lattice the coordinates are given by two integers, $ a(n \hat{x} + m \hat{y})$, and the sub-lattices are characterised by whether $ n+m$ is  even or odd. We have the fermionic fields $ \chi^{\pm} $ and $ \psi^{\pm}$ occupying the points $ n+m$ even and odd respectively. For the matter of representation, we define $ \varphi^{\pm}_{n,m} = (\chi^{\pm}_{n,m} ,  \psi^{\pm}_{n,m} )^{T}$, and indeed $\psi^{\pm}_{n,m} =0 $, $\chi^{\pm}_{n,m} =0 $ for $ n+m$ even and odd respectively. In the bulk we study the following Hamiltonian,

\begin{equation}\label{eq:H2D}
    H_{2D}=a\sum_{n,m}
    \begin{pmatrix}h^{+}_{n,m}&0\\0&h^{-}_{n,m}\end{pmatrix},
\end{equation}

\noindent with

\begin{equation}\label{eq:h2D-def}
\begin{split}
    h^{\pm}_{n,m}
    =\frac{iA}{2a}\bigl[\pm(\varphi_{n,m}^{\pm})^{\dagger}
      \sigma_{x}\varphi_{n+1,m}^{\pm}
     -(\varphi_{n,m}^{\pm})^{\dagger}\sigma_{y}\varphi_{n,m+1}^{\pm}\bigr]
    \\[4pt]
    +\frac{1}{2a^{2}}(\varphi_{n,m}^{\pm})^{\dagger}\sigma_{z}
    \Bigl[B\bigl(\varphi_{n+1,m+1}^{\pm} +\varphi_{n-1,m+1}^{\pm}
                \\ +\varphi_{n+1,m-1}^{\pm}+\varphi_{n-1,m-1}^{\pm}\bigr)\\
    +(2M_{0}a^{2}-4B)\varphi_{n,m}^{\pm}\Bigr]+\text{h.c.},
\end{split}
\end{equation}

where $ \sigma_{i}$ are the Pauli matrices, $ A,B,M_{0}$ are the parameters of TI we are considering in the bulk. When the Fourier transform is applied on $ \varphi^{\pm}_{n,m}$, we get $ p_{x}, p_{y} \in [-\frac{\pi}{a}, \frac{\pi}{a}]$ Fourier variables and since the square lattice is covered by two sublattices, we should have $ \abs{p_{x}+ p_{y}} \leq \frac{\pi}{a}$. Then the Fourier conjugate of $ h^{\pm}$ is written

\begin{equation}
\begin{split}
        \tilde{h}^{\pm}(p_{x}, p_{y}) = -  \frac{A}{a} (\pm  \sin{(p_{x} a) } \sigma_{x}+ \sin{(p_{y} a) } \sigma_{y})\\
    + ( \frac{2 B}{a^{2}}( \cos{(p_{x} a )}\cos{(p_{y} a )}-1 )  + M_{0}  )\sigma_{z},
\end{split}
\end{equation}

\noindent In the continuum limit the bulk energy spectrum becomes

\begin{equation}\label{eq:E2D}
    E^{\pm}_{2D}
    =\pm\sqrt{A^{2}(p_{x}^{2}+p_{y}^{2})
    +(M_{0}-B(p_{x}^{2}+p_{y}^{2}))^{2}},
\end{equation}

\noindent in agreement with the standard BHZ spectrum~\cite{Qi_2011,konig}.

\paragraph*{Topological solitons.}
In condensed matter language, the vacuum is a trivial insulator
with positive mass, while a TI has a negative mass term in part of
the parameter space~\cite{Shen2017-za}.
The existence of edge states requires a sign change of the mass
--- a topological soliton (domain wall)~\cite{Jackiw:1975fn} ---
at the TI-vacuum interface.
For $H_{2D}$, edge states exist when $M_{0}a^{2}<2B$ with $A=1$ and
$B>0$.

\paragraph*{Zero-energy modes.}
Following Ref.~\cite{konig}, we find the zero-energy solutions of
$H_{2D}$ in the continuum.
At $p_{x}=0$ and $A=1$, the operator $h^{\pm}$ reduces to

\begin{equation}\label{eq:hpm-y}
    \tilde{h}_{y}^{\pm}
    =(\tilde{\varphi}^{\pm})^{\dagger}
    \bigl(-i\sigma_{y}\partial_{y}
     +\sigma_{z}(B\partial_{y}^{2}+M_{0})\bigr)
    \tilde{\varphi}^{\pm},
\end{equation}

\noindent where $\tilde{\varphi}^{\pm}(x,y)=\lim_{a\to0}\varphi^{\pm}_{n,m}$.
Treating $\tilde{\varphi}^{\pm}$ as a classical field and applying
the same ansatz as in Ref.~\cite{Jackiw:1975fn}, the classical
equation of motion gives

\begin{equation}\label{eq:eq-of-motion-in-y}
    \bigl(\partial_{y}-\sigma_{x}(B\partial_{y}^{2}+M_{0})\bigr)
    \tilde{\varphi}^{\pm}(x,y)=0.
\end{equation}

\noindent Notably, Eq.~\eqref{eq:eq-of-motion-in-y} does not depend
on the chirality index $\pm$, so the two chiralities share the same
zero-mode profile.
The general solution is

\begin{equation}\label{eq:gen-sol}
    \tilde{\varphi}^{\pm}(x,y)
    =C_{0}e^{\lambda y}\omega_{0}^{\pm}(x)
    +C_{1}e^{-\lambda y}\omega_{1}^{\pm}(x),
\end{equation}

\noindent where $\sigma_{x}\omega_{0,1}^{\pm}=\pm\omega_{0,1}^{\pm}$
(eigenvectors of $\sigma_{x}$ in the flavour basis, analogous to
$\tilde{\psi}_{0,1}$ of the continuum flavoured theory), and
$\lambda$ satisfies

\begin{equation}
    B\lambda^{2}-\lambda+M_{0}=0,
    \qquad
    \lambda_{\pm}=\frac{1\pm\sqrt{1-4BM_{0}}}{2B}.
\end{equation}

\paragraph*{Edge states.}
For $B>0$ and $0<M_{0}\leq B$, we have $\text{Re}(\lambda_{\pm})\geq0$.
With open boundary conditions
$\tilde{\varphi}^{\pm}(x,\pm l_{y})=0$, the normalised solutions
localised at the two boundaries are

\begin{subequations}\label{eq:edge-states}
\begin{align}
    \tilde{\varphi}^{\pm}_{0}(x,y)
    &=C^{+}_{0}\bigl(e^{\lambda_{+}(y-l_{y})}
               -e^{\lambda_{-}(y-l_{y})}\bigr)\omega_{0}^{\pm}(x),
    \\[4pt]
    \tilde{\varphi}^{\pm}_{1}(x,y)
    &=C^{+}_{1}\bigl(e^{-\lambda_{+}(y+l_{y})}
               -e^{-\lambda_{-}(y+l_{y})}\bigr)\omega_{1}^{\pm}(x).
\end{align}
\end{subequations}

\noindent $\tilde{\varphi}^{\pm}_{0}$ is localised at $y=l_{y}$
(top edge) and $\tilde{\varphi}^{\pm}_{1}$ at $y=-l_{y}$
(bottom edge).
With the normalisation $(\omega_{0,1}^{\pm})^{\dagger}\omega_{0,1}^{\pm}=1$
one finds $C^{+}_{0}=C^{+}_{1}$ and, in the large-$l_{y}$ limit,

\begin{equation}
    C^{+}_{0}=C^{+}_{1}
    \simeq\sqrt{\frac{2\lambda_{+}\lambda_{-}(\lambda_{+}+\lambda_{-})}{(\lambda_{+}-\lambda_{-})^{2}}}.
\end{equation}

\paragraph*{Edge Hamiltonians.}
Integrating the bulk Hamiltonian along the $y$-axis and using
$\sigma_{x}\omega_{0,1}^{\pm}=\pm\omega_{0,1}^{\pm}$, the effective
Hamiltonians at the top and bottom edges are

\begin{equation}\label{eq:Hedge}
\begin{split}
    H_{T}
    &=i\int dx\bigl[(\omega^{+}_{0})^{\dagger}\partial_{x}\omega^{+}_{0}
      -(\omega^{-}_{0})^{\dagger}\partial_{x}\omega^{-}_{0}\bigr],
    \\[6pt]
    H_{B}
    &=-i\int dx\bigl[(\omega^{+}_{1})^{\dagger}\partial_{x}\omega^{+}_{1}
      -(\omega^{-}_{1})^{\dagger}\partial_{x}\omega^{-}_{1}\bigr].
\end{split}
\end{equation}

\begin{figure}[t]
    \centering
    \includegraphics[width=0.45\textwidth]{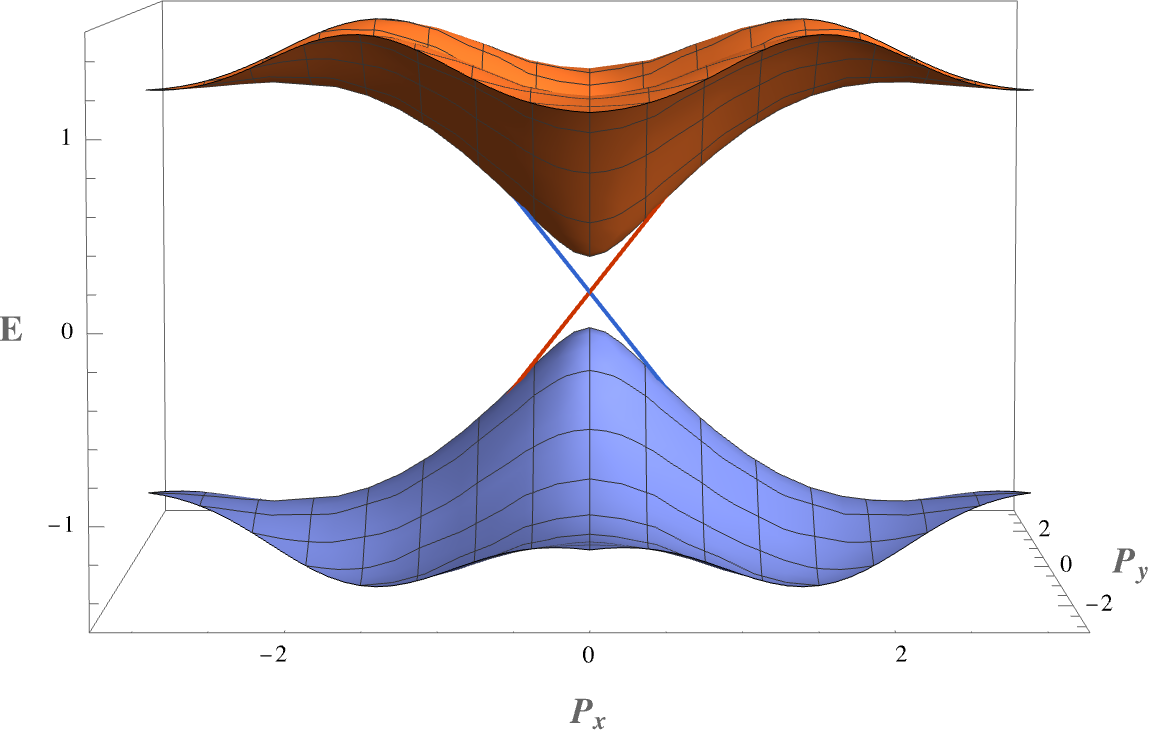}
    \caption{Bulk energy spectrum of $H_{2D}$ (gapped valence and
    conduction bands for $a=1$, $A=1$, $B=0.33$, $M_{0}=0.5$)
    together with the gapless top-edge dispersion.
    Red and blue lines represent chirality $+$ and $-$, respectively.}
    \label{fig:dispersion-relation}
\end{figure}

\noindent Both are free Dirac Hamiltonians: the bulk gap is finite,
but the edge spectrum is gapless.
The bulk and edge dispersions are compared in
Fig.~\ref{fig:dispersion-relation}.

\medskip

The equations of motion of $H_{T}$ for $\omega_{0}^{\pm}$ coincide
with those of $H_{f}^{c}$ for $\tilde{\psi}_{0}^{\pm}$, and those
of $H_{B}$ for $\omega_{1}^{\pm}$ coincide with those for
$\tilde{\psi}_{1}^{\pm}$.
We may therefore identify
$\omega^{\pm}_{0,1}(x)=\tilde{\psi}^{\pm}_{0,1}(x)$,
giving $H_{T}+H_{B}=H_{f}^{c}$. This is the key result of this subsection: the two flavours of the
continuum flavoured theory are spatially separated as helical edge
states on opposite boundaries of the 2+1D TI.
In the broader picture, what we have done is to acknowledge the
doublers as legitimate solutions and provide them with a physical
domain (the opposite boundary of the TI), rather than discarding or
suppressing them. We emphasise that the $\mathbb{Z}_2$ flavour symmetry is not broken by the topological 
insulator embedding; rather, it is \textit{geometrically resolved}. The two flavour sectors 
remain related by $\mathbb{Z}_2$ in the bulk, but are spatially localised on opposite 
boundaries, rendering them physically distinguishable and independently accessible.

\subsection{\label{sec:gauged-TI}Gauged topological insulator embedding}

In this subsection we show that, in contrast with the strictly $(1{+}1)$D model 
\eqref{eq:Hschf-cont}, gauging the $(2{+}1)$D TI embedding of 
Subsec.~\ref{sec:TI-realisation} in a constant background 
electric field produces a boundary theory that factorises into 
\emph{two decoupled} single-flavour Schwinger models, one on each 
edge of the ribbon. The factor of $2$ in Eq.~\eqref{eq:lattice-anomaly} 
is thereby realised geometrically as one quantum of Schwinger anomaly 
per edge.

\paragraph*{Gauged bulk Hamiltonian.}
We couple the bulk Hamiltonian $H_{2D}$ of Eq.~\eqref{eq:h2D-def} 
minimally to a compact $U(1)$ gauge field by attaching parallel 
transporters $\mathcal{U}^{i}_{n,m}=e^{ig\mathcal{A}^{i}_{n,m}a}$ 
on each horizontal ($i=x$) and vertical ($i=y$) bulk link, with 
electric-field operators $\mathcal{L}^{i}_{n,m}$ satisfying 
$[\mathcal{L}^{i}_{n,m},\mathcal{U}^{j}_{n',m'}] = 
\delta^{ij}\delta_{nn'}\delta_{mm'}\mathcal{U}^{j}_{n,m}$; on the 
diagonal links of the $\sigma_{z}$ term, the parallel transport is 
the path-ordered product $\mathcal{U}^{x}_{n,m}\mathcal{U}^{y}_{n+1,m}$. 
For a constant temporal-gauge background $\mathcal{A}^{x}=\mathcal{A}$, 
$\mathcal{A}^{y}=0$, the gauged bulk dispersion is obtained from 
$\tilde h^{\pm}(p_{x},p_{y})$ by the rigid shift $p_{x}\to p_{x}-g\mathcal{A}$. 
The edge-existence condition $M_{0}a^{2}<2B$ is unaffected and the 
bulk remains gapped.

\paragraph*{Edge Hamiltonians.}
The transverse zero-mode operator $\tilde h^{\pm}_{y}$ in 
Eq.~\eqref{eq:hpm-y} depends only on $p_{y}$ and on the bulk mass, 
so the constant background leaves the localisation profiles 
$\tilde\varphi_{0}^{\pm}(x,y)$ at $y=l_{y}$ and 
$\tilde\varphi_{1}^{\pm}(x,y)$ at $y=-l_{y}$ of 
Eq.~\eqref{eq:edge-states} exactly intact. Integrating $H_{2D}^{g}$ 
against these profiles and defining the projected edge gauge fields 
$\mathcal{A}_{T,B}(x)\equiv\lim_{a\to0}\mathcal{A}^{x}(x,\pm l_{y})$ 
with conjugate electric fields 
$\mathcal{E}_{T,B}(x)=g\mathcal{L}^{x}(x,\pm l_{y})$ gives
\begin{equation}\label{eq:Hedge-gauged}
\begin{aligned}
   H_{T}^{g} &= i\!\int\! dx\,\bigl[(\omega_{0}^{+})^{\dagger}D_{T}\omega_{0}^{+}
                 -(\omega_{0}^{-})^{\dagger}D_{T}\omega_{0}^{-}\bigr]
              + \tfrac{1}{2}\!\!\int\! dx\,\mathcal{E}_{T}^{2},\\
   H_{B}^{g} &= -i\!\int\! dx\,\bigl[(\omega_{1}^{+})^{\dagger}D_{B}\omega_{1}^{+}
                  -(\omega_{1}^{-})^{\dagger}D_{B}\omega_{1}^{-}\bigr]
              + \tfrac{1}{2}\!\!\int\! dx\,\mathcal{E}_{B}^{2},
\end{aligned}
\end{equation}
with $D_{T,B}\equiv\partial_{x}-ig\mathcal{A}_{T,B}$.

\paragraph*{Spatial decoupling and factorisation.}
Two independent mechanisms ensure that the top and bottom edges decouple.
First, on the matter side, the cross-edge overlap of the localisation
profiles is exponentially suppressed,
$\int dy\, |\tilde\varphi_0|^2 |\tilde\varphi_1|^2 \sim e^{-4\,\mathrm{Re}(\lambda_\pm)\, l_y}$,
so that no fermionic tunnelling process connects the two edges in the
wide-ribbon limit. Second, on the gauge side, the background is purely
longitudinal ($A_x = \mathcal{A}$, $A_y = 0$), so that no transverse
electric field is excited in the bulk; since $\mathcal{E}_T$ and
$\mathcal{E}_B$ are each built solely from the longitudinal component
$L^x(x,\pm l_y)$ of the gauge field at their respective edge, there is
no transverse electric field to correlate the two rows $y = +l_y$ and
$y = -l_y$, either dynamically or through the Gauss constraint.
Consequently, $\mathcal{A}_T$ and $\mathcal{A}_B$ are independent
dynamical variables in the wide-ribbon limit, and the gauged
TI-embedded flavoured Schwinger model factorises as
\begin{equation}\label{eq:TI-factorisation}
   H_{T}^{g}+H_{B}^{g}\;\xrightarrow[l_{y}\to\infty]{}\;
   H_{\mathrm{sch}}^{(\alpha=0)}\,\oplus\,H_{\mathrm{sch}}^{(\alpha=1)},
\end{equation}
i.e.\ two genuinely decoupled single-flavour massless Schwinger 
models, one on each edge. This contrasts sharply with the strictly 
$(1{+}1)$D theory Eq.~\eqref{eq:Hschf-cont}, where the two flavours 
share a common photon $\tilde A(x)$ and the continuum theory is the 
two-flavour Schwinger model with iso-singlet photon mass 
$m^{2}=2g^{2}/\pi$; in the TI realisation each edge supports an 
independent photon of mass $g^{2}/\pi$ at two spatially separated 
locations.

\section{\label{sec:prospects} Conclusion}

\paragraph*{Summary.}
We have introduced and systematically studied the flavoured lattice 
Schwinger model, a $(1{+}1)$-dimensional $U(1)$ lattice gauge theory 
obtained by applying the flavoured fermion 
construction~\cite{bakircioglu-fd-qca} to the standard lattice 
Schwinger model. The construction staggers a $\mathbb{Z}_{2}$ 
flavour degree of freedom across even and odd sites, preserving 
\emph{both} vector and axial $U(1)$ symmetries at the lattice level, 
in contrast to staggered fermions which break axial symmetry 
explicitly. The continuum limit is the \emph{two-flavour} massless 
Schwinger model: the flavours $\alpha\in\{0,1\}$ share a single 
dynamical $U(1)$ gauge field, and the Schwinger mechanism generates 
a single iso-singlet boson of mass $m^{2}=2g^{2}/\pi$. 
Section~\ref{sec:chiral anomaly} derives the chirally anomalous response 
and bosonisation: Subsec.~\ref{sec:lattice-anomaly} gives the lattice 
chiral anomaly equation \eqref{eq:lattice-anomaly}; unlike in staggered 
or Wilson formulations, $Q_{G}^{A}$ is a finite, well-defined, 
gauge-invariant lattice observable whose non-conservation arises 
dynamically through the electric-field term.

\paragraph*{Prospects.}
Building on the exact, gauge-invariant lattice axial charge $Q^A_G$ established here, several natural extensions present themselves.

\emph{Non-Abelian gauge theories.}
The flavoured fermion construction generalises naturally to $SU(N)$ gauge theories. Pursuing this extension provides a direct pathway to compute the non-Abelian chiral anomaly, offering a new framework to investigate the intricate interplay between flavour, colour, and the Nielsen--Ninomiya theorem on the lattice.

\emph{Quantum simulation.}
Reinterpreting doubler modes as physical flavour degrees of freedom is highly advantageous for quantum simulation architectures, where additional internal states can be engineered explicitly~\cite{PhysRevX.10.021041,Martinez2016}. In such platforms, $Q_{G}^{A}$ becomes a directly measurable observable, making the dynamical generation of the anomaly experimentally accessible.

\emph{Gauged non-Abelian sector.}
Gauging the emergent $SU_{V}(2)$ exposed by Subsec.~\ref{sec:WZW} would 
produce a gauged $SU(2)_{1}$ WZW model~\cite{Affleck:1985wb}, a 
$(1{+}1)$D non-Abelian generalisation of the Schwinger model whose 
lattice realisation within the flavoured construction is left for 
future work.

\section{\label{sec:acknowledgements} Acknowledgements}
I am profoundly grateful to Yuya Tanizaki for his detailed questions, which were instrumental in resolving an earlier inaccuracy and motivated the connection to the Wess--Zumino--Witten model. I also wish to extend my sincere thanks to Pablo Arrighi for his continuous support, insightful suggestions, and assistance in refining the abstract. Finally, I gratefully acknowledge Pablo Arnault for his encouragement and constructive feedback regarding the structure of this manuscript.

\bibliographystyle{apsrev4-1}
\bibliography{ref}

\appendix

\section{Derivation of the Fermionic Correlators}
\label{app:correlators}

We derive the equal-time ground-state correlators used in
Sec.~\ref{sec:chiral anomaly}.
The calculation proceeds in three steps:
(A.1) free theory without chirality;
(A.2) the gauged extension and gauge-independence;
(A.3) the full chiral case.

\subsection*{A.1\quad Free theory}

We work with the free flavoured Hamiltonian ($A_{n}=0$), labelling
unit cells by $m$ (even site $n=2m$, odd site $n{+}1=2m{+}1$).
For a single chirality $\sigma=+$ the relevant hopping term is

\begin{equation}
    H_{f}^{+}=\frac{i}{2a}\sum_{m}
    \bigl[(\chi^{+}_{2m})^{\dagger}
    (\psi^{+}_{2m+1}-\psi^{+}_{2m-1})\bigr]+\text{h.c.}
\end{equation}

The unit cell has size $2a$, so the BZ is $k\in[-\pi/2a,\pi/2a]$.
Expanding in Fourier modes,

\begin{equation}\label{eq:app_FT}
    \chi^{+}_{2m}=\sqrt{\tfrac{2}{N}}\sum_{k}e^{i(2m)ka}\,c^{+}_{k},
    \hspace{1mm}
    \psi^{+}_{2m+1}=\sqrt{\tfrac{2}{N}}\sum_{k}e^{i(2m+1)ka}\,d^{+}_{k},
\end{equation}

\noindent where $c_{k}^{\pm},d_{k}^{\pm}$ satisfy canonical
anti-commutation relations.
Substituting,

\begin{equation}\label{eq:app_Hk}
    H_{f}^{+}=-\frac{1}{a}\sum_{k}\sin(ka)
    \bigl(c^{+\dagger}_{k}d^{+}_{k}+d^{+\dagger}_{k}c^{+}_{k}\bigr).
\end{equation}

At each $k$ the $2\times2$ single-particle matrix has eigenvalues
$\pm|\sin(ka)|/a$.
The lower-energy (filled) mode and its expectation value are

\begin{equation}\label{eq:app_occ}
    \alpha^{+}_{k}=\frac{c^{+}_{k}+\mathrm{sgn}(\sin ka)\,d^{+}_{k}}{\sqrt{2}},
    \qquad
    \langle c^{+\dagger}_{k}d^{+}_{k}\rangle
    =\tfrac{1}{2}\,\mathrm{sgn}(\sin ka).
\end{equation}

By translational invariance the real-space correlator is
site-independent:

\begin{equation}\label{eq:app_int}
    \langle(\chi^{+}_{n})^{\dagger}\psi^{+}_{n+1}\rangle
    \xrightarrow{N\to\infty}
    \frac{1}{2\pi}\int_{-\pi/2}^{\pi/2}
    e^{iu}\,\mathrm{sgn}(\sin u)\,du,
\end{equation}

\noindent where $u=ka$.
Since $\mathrm{sgn}(\sin u)=\mathrm{sgn}(u)$ for $u\in(-\pi/2,\pi/2)$:

\begin{align}
    \langle(\chi^{+}_{n})^{\dagger}\psi^{+}_{n+1}\rangle
    &=\frac{1}{2\pi}\!\left[\int_{0}^{\pi/2}\!e^{iu}\,du
       -\int_{-\pi/2}^{0}\!e^{iu}\,du\right]
    \nonumber\\
    &=\frac{1}{2\pi i}
    \bigl[(e^{i\pi/2}-1)-(1-e^{-i\pi/2})\bigr]
    =\frac{i}{\pi}.
    \label{eq:app_result_free}
\end{align}

Since $(\chi^{+}_{n})^{\dagger}\psi^{+}_{n+1}
-(\psi^{+}_{n+1})^{\dagger}\chi^{+}_{n}$ is anti-Hermitian,
$\langle(\psi^{+}_{n+1})^{\dagger}\chi^{+}_{n}\rangle=-i/\pi$, and

\begin{equation}\label{eq:app_corr_plus_free}
    \bigl\langle(\chi^{+}_{n})^{\dagger}\psi^{+}_{n+1}
    -(\psi^{+}_{n+1})^{\dagger}\chi^{+}_{n}\bigr\rangle
    =\frac{2i}{\pi}.
\end{equation}

\subsection*{A.2\quad Gauged theory: gauge-independence}

For the gauged Hamiltonian $H_{\text{sch}}^{f}$ with a uniform
$U(1)$ background $ \forall U_{n}=e^{i\theta}, \, \, \theta = g A a$ on all links, the Fourier
analysis of Eq.~\eqref{eq:app_Hk} goes through with the replacement
$e^{\pm ika}\to e^{\pm i(ka-\theta)}$, giving

\begin{equation}
    H_{\text{sch}}^{f,+}\big|_{U_{n}=e^{i\theta}}
    =-\frac{1}{a}\sum_{k}\sin(ka-\theta)
    \bigl(c^{+\dagger}_{k}d^{+}_{k}+d^{+\dagger}_{k}c^{+}_{k}\bigr).
\end{equation}

The dispersion is shifted $k\to k-\theta/a$, and the
gauge-invariant correlator evaluates to

\begin{equation}
    \langle(\chi^{+}_{n})^{\dagger}U_{n}^{\dagger}\psi^{+}_{n+1}\rangle
    =\frac{e^{-i\theta}}{2\pi}\int_{-\pi/2}^{\pi/2}
    e^{iu}\,\mathrm{sgn}(\sin(u-\theta))\,du.
\end{equation}

Substituting $v=u-\theta$ and splitting at $v=0$ (which lies inside
the integration range for $|\theta|<\pi/2$):

\begin{align}
    \langle(\chi^{+}_{n})^{\dagger}U_{n}^{\dagger}\psi^{+}_{n+1}\rangle
    &=\frac{1}{2\pi i}\Bigl[
       e^{-i\theta}\underbrace{(e^{i\pi/2}+e^{-i\pi/2})}_{=0}-2
    \Bigr]=\frac{i}{\pi}.
\end{align}

The result is independent of $\theta$.
This is not accidental: in one spatial dimension there are no
gauge-invariant plaquettes, the gauge field is pure gauge, and one
can always reach the axial gauge $U_{n}=1$ by a local
transformation~\cite{hamiltonianlgt,Smit_2002}.
Hence the gauge-invariant observable
$\langle(\chi^{+}_{n})^{\dagger}U_{n}^{\dagger}\psi^{+}_{n+1}
-(\psi^{+}_{n+1})^{\dagger}U_{n}\chi^{+}_{n}\rangle$
must equal its free-theory value:

\begin{equation}\label{eq:app_corr_plus_gauged}
    \bigl\langle(\chi^{+}_{n})^{\dagger}U_{n}^{\dagger}\psi^{+}_{n+1}
    -(\psi^{+}_{n+1})^{\dagger}U_{n}\chi^{+}_{n}\bigr\rangle
    =\frac{2i}{\pi}.
\end{equation}

\subsection*{A.3\quad Including chirality}

For $\sigma=-$ the Hamiltonian carries a global sign flip relative
to $\sigma=+$:

\begin{equation}
    H_{f}^{-}=-\frac{i}{2a}\sum_{m}
    \bigl[(\chi^{-}_{2m})^{\dagger}
    (\psi^{-}_{2m+1}-\psi^{-}_{2m-1})\bigr]+\text{h.c.},
\end{equation}

\noindent giving in momentum space
$H_{f}^{-}=+(1/a)\sum_{k}\sin(ka)(c^{-\dagger}_{k}d^{-}_{k}+\text{h.c.})$.
The sign flip swaps which linear combination is the lower-energy
mode:

\begin{equation}
    \langle c^{-\dagger}_{k}d^{-}_{k}\rangle
    =-\tfrac{1}{2}\,\mathrm{sgn}(\sin ka).
\end{equation}

The BZ integral then picks up a global minus sign, giving
$\langle(\chi^{-}_{n})^{\dagger}\psi^{-}_{n+1}\rangle=-i/\pi$.
Collecting both chiralities, and noting that the
gauge-independence argument of Sec.~A.2 applies unchanged for
$\sigma=-$:

\begin{equation}\label{eq:app_final}
    \bigl\langle(\chi^{\pm}_{n})^{\dagger}U_{n}^{\dagger}\psi^{\pm}_{n+1}
    -(\psi^{\pm}_{n+1})^{\dagger}U_{n}\chi^{\pm}_{n}\bigr\rangle
    =\pm\frac{2i}{\pi}.
\end{equation}

This result is purely imaginary (as required by anti-Hermiticity),
independent of the lattice site $n$ (translational invariance), and
independent of the gauge background ($1{+}1$D pure gauge).
It is the key input to the chiral anomaly calculation in
Sec.~\ref{sec:chiral anomaly}.

\end{document}